\documentclass[12pt]{iopart}
\usepackage{amssymb}
\usepackage{graphicx}
\begin{document}

\title[Population explosion suppressed by noise]{Population explosion suppressed by noise: 
Stationary distributions and how to simulate them}

\author{P F G\'ora}

\address{Marian Smoluchowski Institute of Physics and 
Mark Kac Complex Systems Research Centre, 
Jagellonian University, Reymonta~4, 30--059 Krak\'ow, Poland}

\ead{gora@if.uj.edu.pl}

\begin{abstract}
We show that two dynamical systems exhibiting
very different deterministic behaviours possess very similar stationary
distributions when stabilized by a multiplicative Gaussian white
noise. We also discuss practical aspects of numerically simulating
these systems. We show that there exists a noise level
that is optimal in the sense that the interval during which 
discrete-time versions of the systems remain physical is maximized.
Analytical results are illustrated by numerical examples.
\end{abstract}


\submitto{\NJP}

\maketitle

\section{Introduction}

Population dynamics is one of the natural areas of application of the theory of
stochastic processes: Parameters of deterministic models are perturbed by random
fluctuations in order to model the influence of an ever-changing environment 
consisting of a multitude of individual ``agents''. Analytical solutions are
known for only a handful of such systems. A majority of physically important 
systems is accessible only via numerical simulations. It is, therefore, 
particularly important to thoroughly understand those few whose analytical
properties are known and test how well the existing numerical methods can
approximate them.  When a rigorous mathematical statement can be made about 
a system described by a stochastic differential
equation (SDE), one expects that it will be corroborated by numerical simulations or by
an actual physical experiment, if the latter is possible: after all, an SDE 
is a mathematical idealization of a physical reality that is discrete by its very nature. 
If the results of the two approaches, continuous time SDE vs.\ discrete time numerical
simulations, do not agree, there are reasons for a serious concern.

One of the most surprising results, reported recently by Mao and co-workers in
Ref.~\cite{marion}, is the fact that a multiplicative noise can stabilize 
a~system whose deterministic counterpart is divergent. The class of systems 
discussed in that Reference may comprise several interacting species but its
simplest form, to which we will restrict ourselves within this paper, is described
by a scalar Langevin equation

\begin{equation}\label{marion:eksplozja}
\dot x = rx(1\pm x) + \sigma x^2\xi(t)\,,
\end{equation}

\noindent where $r>0$ and $\xi(t)$ is a Gaussian white noise (GWN),
interpreted in the sense of Ito, with 
$\left\langle\xi(t)\right\rangle=0$ and 
$\left\langle\xi(t_1)\xi(t_2)\right\rangle=\delta(t_1-t_2)$. If there is 
no noise, $\sigma=0$, the ``$-$'' sign correspond to the well-known logistic
equation, while the ``$+$'' sign describes a ``population'' that diverges in a finite
time. This latter case defies one of the ecological laws that a population should
be self-limited~\cite{oikos}, although Eq.~(\ref{marion:eksplozja})
perhaps can be used to model explosive phenomena observed in some chemical,
nuclear or stellar systems where a~divergence is interpreted as a~destruction of
the original system. However, our primary purpose of discussing the ``$+$'' sign
is to present the suprising effects that noise can have on a~deterministically 
divergent system.

It is rigorously shown in Ref.~\cite{marion} that for any nonzero noise, 
$\sigma\not=0$, if the system~(\ref{marion:eksplozja}) starts from a positive
initial value, it will for all times remain positive and bounded almost surely, regardless
of the choice of the sign inside the brackets.
This means that even a~tiny addition of the noise can stabilize the system.
We note that a proof of this statement runs in two stages:
First it is proven that the system remains positive, and later this fact is used to
prove that the system remains bounded. 

Apart from the formal proof, Ref.~\cite{marion} provides several numerical examples.
These examples, however, are much less convincing than the fine mathematical
points they are supposed to illustrate. First, it is not clear whether the numerical
trajectories presented are typical or handpicked ``best'' ones. Second, the
noise has been approximated by an uncorrelated Bernoulli noise while it is believed 
that one should either use an exponentially correlated dichotomic noise and then take 
a certain double limit to simulate a GWN \cite{chris}, or directly use good generators 
of a~GWN. We will see that the problem of simulating the system described by 
the equation (\ref{marion:eksplozja}) is unexpectedly tricky: Even though this system almost
surely never diverges, we will show that its discretizations realized by several
popular and well-established methods do diverge.

This paper is organized as follows: In Section~\ref{distribution} we construct stationary
probability distributions corresponding to the equation (\ref{marion:eksplozja}). 
Several interacting
species may not converge to a stable fixed point even if their trajectories are
bounded. They may display oscillations instead and may not have a stationary distribution. 
However, if there is only one species, finding its stationary
distribution is a~straightforward task. 
We will show that systems whose deterministic dynamics are completely different may
possess similar stationary distribution when perturbed by the noise. Then in 
Sections~\ref{numerics} and~\ref{higher} we will discuss how to simulate the system 
(\ref{marion:eksplozja}). 
In particular, we will show that there is an ``optimal'' noise level that maximizes 
the average lifespan of a discretized system. Conclusions are given in 
Section~\ref{conclusions}.

\section{The stationary distribution}\label{distribution}

The equation (\ref{marion:eksplozja}) leads to the following Fokker-Planck equation:

\begin{equation}\label{marion:Fokker}
\frac{\partial P(x,t)}{\partial t} =
-r\frac{\partial}{\partial x}\left[x(1\pm x)P(x,t)\right]
+\frac{\sigma^2}{2}\frac{\partial^2}{\partial x^2}
\left[x^4P(x,t)\right].
\end{equation}

\noindent A stationary solution to this equation can be easily found by
standard methods \cite{Risken,Gardiner}. It is, however, instructive to
take a different approach. We make a substitution

\begin{equation}\label{marion:substitution}
y=\frac{1}{x}\,.
\end{equation}

\noindent Note that because $x$ is nonnegative almost surely,
the substitution (\ref{marion:substitution}) is always valid,
i.e.\ we do not risk a~division by zero almost surely. As a result of this substitution,
the Fokker-Planck equation (\ref{marion:Fokker}) in the Ito interpretation
is transformed into~\cite{vankampen} 

\begin{equation}\label{marion:Fokker-y}
\frac{\partial P(y,t)}{\partial t} = 
\frac{\partial}{\partial y}\left[\left(ry \pm r -\frac{\sigma^2}{y}\right)P(y,t)\right]
+\frac{\sigma^2}{2}\frac{\partial^2 P(y,t)}{\partial y^2}\,,
\end{equation}

\noindent which, in turn, corresponds to the following Langevin equation:

\begin{equation}\label{marion:Langevin-y}
\dot y = -\left(ry \pm r -\frac{\sigma^2}{y}\right) + \sigma\xi(t)\,.
\end{equation}

\noindent The deterministic part of this equation describes an overdamped motion in 
a ``potential''

\begin{equation}\label{marion:potential}
U(y) = \frac{1}{2}ry^2 \pm ry-\sigma^2\ln y\,.
\end{equation}

\noindent We can see that the noise interpreted in the sense of Ito introduces 
a barrier preventing $y$ from 
crossing zero, or preventing the population $x$ from diverging, as predicted
in Ref.~\cite{marion}. Note that this barrier is absent if the equation (\ref{marion:eksplozja})
is interpreted in the sense of Stratonovich, and indeed there is no proof
that the population $x$ does not explode in this case. 
Finding the stationary solution to 
the equation (\ref{marion:Fokker-y}) is now a matter of a simple calculation:

\begin{equation}\label{marion:stationary-y}
P_{\mathrm{st}}(y) = {\cal N}y^2\exp\left[-\frac{r}{\sigma^2}y(y\pm2)\right],
\end{equation}

\noindent or after transforming back to the original variable,

\begin{equation}\label{marion:stationary-x}
P_{\mathrm{st}}(x) = \frac{\cal N}{x^4}\exp\left[-\tilde r\,\frac{1\pm2x}{x^2}\right],
\end{equation}

\noindent where $\tilde r=r/\sigma^2$ and the normalization constant equals

\begin{equation}\label{marion:N}
{\cal N}=\frac{4{\tilde r}^{3/2}}
{\sqrt{\pi}\,\exp(\tilde r)(1+2\tilde r)\left(1\mp\mathrm{Erf}\left(\sqrt{\tilde r}\right)\right) 
\mp 2\sqrt{\tilde r}}
\end{equation}

\noindent and $\mathrm{Erf}(\cdot)$ is the error function. It is easy to verify that for
any positive value of $\tilde r$, and for both choices of the sign,
the normalization constant $\cal N$ is positive and finite.
The distribution (\ref{marion:stationary-x}) is the stationary probability distribution 
corresponding to the equation (\ref{marion:Fokker}) with the constraint $x>0$. It 
reaches a maximum at $x=(\sqrt{\tilde r(\tilde r+8)}\pm \tilde r)/4$.

Observe that the ``$-$'' sign in (\ref{marion:eksplozja}) and the subsequent equations
corresponds to a~noisy logistic equation; the corresponding deterministic system is 
certainly stable and never diverges. The sign ``$+$'' corresponds to a system whose
deterministic counterpart diverges very rapidly. However, the choice of this sign
has very little effect on the shape and properties of the distribution
(\ref{marion:stationary-x}). It is indeed very surprising that two systems whose
deterministic behaviours differ so dramatically have very similar stationary
distributions when driven by a GWN.

\section{Numerical simulations}\label{numerics}

When one deals with a Langevin equation in the form

\begin{equation}\label{marion:Langevin}
\dot x = f(x) + g(x)\xi(t)
\end{equation}

\noindent and does not know a stationary distribution corresponding to it, the
usual way to proceed is to use an appropriate discretization in time, solve
such system numerically, let it ``equilibrate'' for a certain time and try 
to reconstruct the (unknown) stationary distribution from the numerical trajectories.
If $\xi(t)$ is a GWN, the celebrated Euler-Maruyama scheme is most frequently used 
for this purpose. In this scheme the numerical trajectory is generated by

\begin{equation}\label{marion:Euler}
x_{n+1} = x_n + hf(x_n) + \sqrt{h}\,g(x_n)\eta_n\,,
\end{equation}

\noindent where $h$ is the time step and $\eta_n$ is a discrete Gaussian white
noise: $\left\langle\eta_n\right\rangle=0$, $\left\langle\eta_n\eta_m\right\rangle
=\delta_{nm}$. The Euler-Maruyama scheme is consistent with the equation (\ref{marion:Langevin}) 
in the Ito interpretation in the limit $h\to0^+$. This scheme is numerically very
cheap but its strong order of convergence is rather small, $\gamma=1/2$~\cite{Kloeden}.

Simulating the noisy logistic equation, corresponding to a~``$-$'' in 
the equation (\ref{marion:eksplozja}), does not pose any particular problem. We will,
therefore, focus on simulating the system whose deterministic counterpart
diverges. If we apply the Euler-Maruyama scheme to the equation (\ref{marion:eksplozja}),
we can see that the population is prevented from exploding by sudden jumps
down. The amplitude of these jumps is scaled by the square of the current value of the
population: the higher the system climbs, the lower it can drop. Large jumps
up are equally possible, but a long series of jumps up is unlikely.
If $x_n$ is large, even a small (and positive) value of $\eta_n$ introduces
a~significant increase in the population size. In the next step, however, 
the probabilities of jumping up and down are equal and even a smaller to the absolute
value but negative $\eta_{n+1}$ ($\eta_{n+1}<0$, $|\eta_{n+1}|<\eta_n$)
can undo the cumulative effect of several jumps up as it is scaled by a
much larger factor.
We know that the ideal, continuous time system neither diverges, nor
becomes negative, but the discretization scheme (\ref{marion:Euler}) applied to
the equation (\ref{marion:eksplozja}) may not obey these principles. 
If the current magnitude of the population becomes too large \textit{or}
negative, the model (\ref{marion:eksplozja}) clearly looses any physical
meaning. We say that the numerical trajectory becomes unphysical
either if $x_{n+1}<0$ or if $x_{n+1}>X\gg0$ and stop the simulation when 
either of these situations  occurs. Observe that the sequence $\{x_n\}$
does not converge to zero without becoming negative: The probability

\begin{equation}\label{marion:refereeB}
\mathrm{Prob}\left(0<x_{n+1}<x_n\,\vert\,x_n>0\right)
=
\mathrm{Prob}\left(\eta_n<-(\sqrt{h}r/\sigma)(1+1/x_n)\right),
\end{equation}

\noindent becomes negligible for $0<x_n\ll1$. On the other hand, 
the probability that $x_{n+1}$ becomes negative equals

\begin{eqnarray}
\mathrm{Prob}\left(x_{n+1}{<}0\,\vert\,x_n{>}0\right) = 
\mathrm{Prob}\left(\eta_n<-\frac{1{+}hr{+}hrx_n}{\sqrt{h}\,\sigma x_n}\right)
\nonumber\\
\label{marion:negative}
{}=\frac{1}{2}\left(1-\mathrm{Erf}\left(\frac{1+hr+hrx_n}{\sqrt{h}\,\sigma x_n}\right)\right).
\end{eqnarray}

\noindent This probability goes to zero in the limit $h\to0^+$, which is nothing more
than a~consistency check of the Euler-Maruyama scheme, but is nonzero for any
$h>0$ and any $x_n>0$. This means that after a certain, perhaps very large but finite,
number of iterations, the Euler-Maruyama scheme \textit{will} produce a negative, i.e.\
unphysical, value. Thus, the discretized system has a finite lifespan, i.e. the time after
which the system, when started from a positive initial value, either becomes negative
or exceedingly large.

Observe that if $0<x_n\ll1$, probability (\ref{marion:negative}) is very small and
increases with an increasing value of $x_n$. If $x_n\gg1$, probability (\ref{marion:negative})
reaches a constant value, $\mathrm{Prob}\left(x_{n+1}{<}0\,\vert\,x_n{>}0\right) 
\simeq(1-\mathrm{Erf}(\sqrt{h}\,r/\sigma))/2$, which may be quite large; for example,
for $h=2^{-16}$ and  $r/\sigma=1$, $\mathrm{Prob}\left(x_{n+1}{<}0\,\vert\,x_n{\gg}1\right) 
\simeq 0.498$. We can see that it is easier for
a discretized version of system (\ref{marion:eksplozja}) to cross zero if the current
value of the process, or the current population, is large than when it is small.
This conclusion is well confirmed by numerical simulations.

Events in which $x_n$ becomes larger than the threshold value $X\gg0$ are very rare. 
If $x_n$
becomes large, it is more likely for it to decrease in a sudden jump and start
its way up again, or even become negative when the simulation stops than gradually
climb up towards the threshold.

If the noise intensity, $\sigma$, is very small, the ``stopping effect'' of the noise
is limited: the population can build up beyond the threshold or, more likely,
to such a value where the probability of the next iterate becoming negative is
significant despite the small value of $\sigma$. On the other hand, in the
limit $\sigma\to\infty$, the probability (\ref{marion:negative}) goes to $1/2$ and the
system becomes unphysical very rapidly. Therefore, we expect the lifespan of a system with
a very small or very large noise intensity to be short. There should be an ``optimal''
range of the intensity of the noise in which the noise effectively stops the
system from climbing too high and at the same time the probability (\ref{marion:negative})
has a modest value. Lifespans of the system with such a noise intensity should be 
larger than for either small or large noise intensities.

\begin{figure}
\begin{center}
\includegraphics{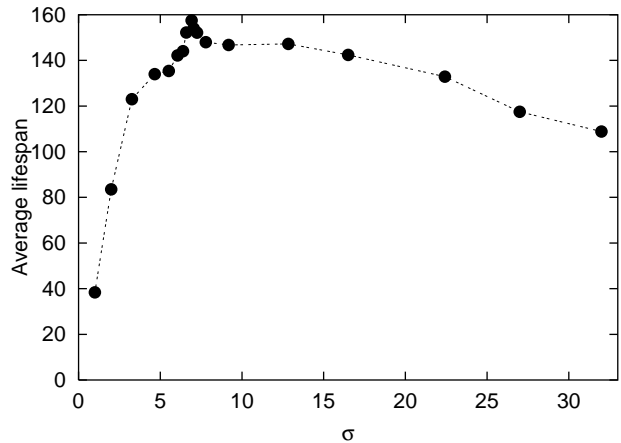}
\end{center}
\caption{Average lifespan of system~(\ref{marion:eksplozja}) discretized by the
Euler-Maruyama scheme (\ref{marion:Euler}) as a function of the noise amplitude,
$\sigma$. The lifespan and $\sigma$ are dimensionless. The line connecting the
points is meant as a guide for the eye only.}
\label{marion:fig-m1}
\end{figure}

We have performed numerical simulations to verify this. We have generated trajectories 
of the equation (\ref{marion:eksplozja}) according to Euler-Maruyama scheme with a time step
$h=2^{-16}\simeq1.5\cdot10^{-5}$. The noise has been generated by Marsaglia algorithm
\cite{Marsaglia} with Mersenne Twister as the underlying uniform generator \cite{Mersenne}.
The simulations were stopped when $x_{n+1}$ became negative or larger than
$X=10^{20}$; the latter events were very rare. For each value of $\sigma$, the lifespans 
were averaged over 1024 realizations
of the process. Results are presented on Figure~\ref{marion:fig-m1}. As we can see, the
average lifespan increases rapidly as $\sigma$ increases, reaches a plateau, and then
gradually decreases. These results show clearly that there is a range of ``optimal''
noise intensities that maximize the lifespan of the discretized process.
A precise location of the maximum is not particularly important
as it may depend on realizations of the noise. 

The average lifespan displays a strong dependence on the time step used to perform
the simulations: The average lifespans increase (decrease) with a decreasing 
(increasing) value of the time step,~$h$, and this increase (decrease) can be
quite significant.
However, with very short time steps, the time needed to perform any realistic
simulations gets prohibitively large.

\begin{figure*}
\begin{center}
\includegraphics{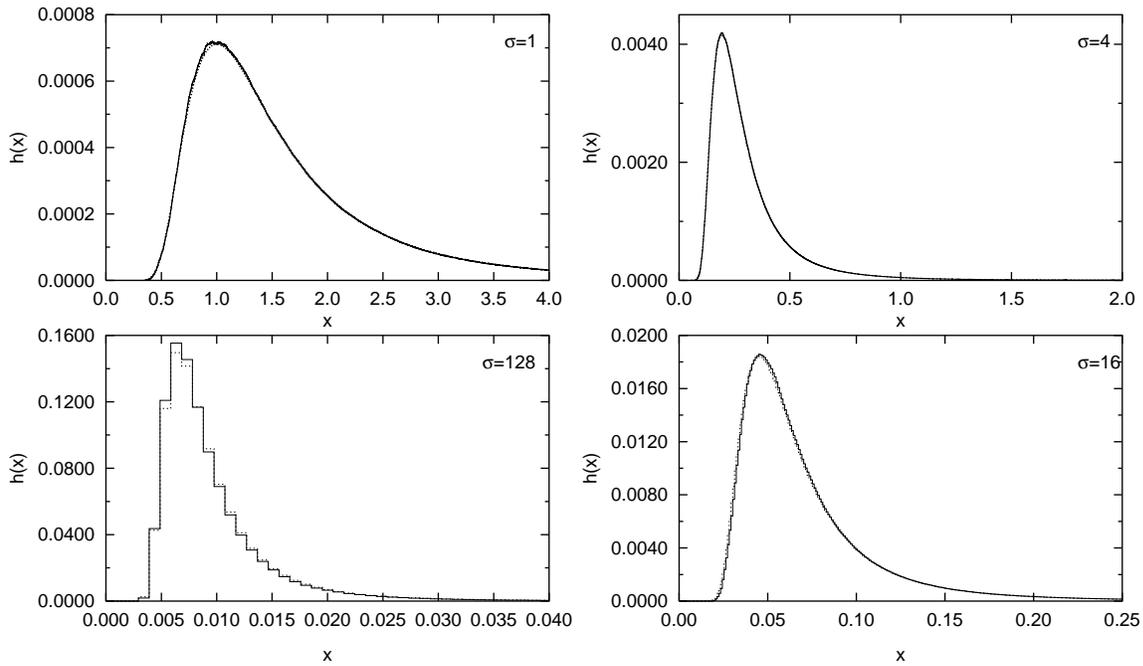}
\end{center}
\caption{Numerical (solid line) and theoretical (broken line) histograms of
distribution (\ref{marion:stationary-x}) corresponding to the deterministically
divergent case (the ``$+$'' sign). Clockwise from top-left $\sigma=1$, 2, 16, and 128,
respectively. The agreement between theoretical and numerical histograms is very good.}
\label{marion:fig-histo}
\end{figure*}

Data gathered during the simulations were also used to draw ``experimental'' histograms
of the distribution (\ref{marion:stationary-x}). Specifically, from all realizations with
$h=2^{-16}$ and
lifespans equal or greater than $64$, values of $x(t)$ for $32\leqslant t\leqslant64$
were collected, binned in intervals of the size $2^{-10}$, averaged over the respective
number of realizations and compared with histograms obtained from the theoretical
distribution. Selected results are presented on Figure~\ref{marion:fig-histo}.
The agreement between theory and numerical experiment is very good for 
$1/2\lesssim\sigma\lesssim1024$, meaning that the Euler-Maruyama scheme reproduces
properties of the equation (\ref{marion:eksplozja}) well.
Results outside this interval are worse because few sufficiently long trajectories
were available and the statistics was poor; for large $\sigma$ one should also use 
a smaller bin size.

\section{Higher order schemes}\label{higher}

Sometimes there is a need to use a higher order discretization scheme. Several such
methods are known in literature and in this Section we will discuss performance
of two of them. One is the Heun method \cite{Mannella} which many people, including
the present author, have applied on several occasions with a great success. However, 
when this method is used to discretize the equation (\ref{marion:eksplozja}), it leads
to negative values of the population very rapidly. The reason for this apparent
failure is the fact that the Heun method is consistent with the equation (\ref{marion:Langevin})
in the Stratonovich, not Ito, interpretation.

The other method, perhaps less popular among physicists, is the Milstein 
scheme~\cite{Milstein}:

\begin{equation}\label{marion:Milstein}
x_{n+1} = x_n + hf(x_n) + \sqrt{h}\,g(x_n)\eta_n + 
\frac{h}{2}g(x_n)g^\prime(x_n)(\eta_n^2{-}1)\,,
\end{equation}

\noindent where $h$, $\eta_n$ are as in the equation (\ref{marion:Euler}) and
$g^\prime(x_n)=(dg/dx)|_{x=x_n}$.

The Milstein scheme differs from Euler-Maruyama only in the last term
and this term vanishes if the noise is purely additive ($g=\mathrm{const}$).
Its computational burden depends on how difficult it is to evaluate the
derivative. In case of a simple polynomial the overall cost does not exceed
twice that for the Euler-Maruyama method.

When this method is applied to the equation (\ref{marion:eksplozja}), we find 
for the probability of the next iterate becoming negative

\numparts
\begin{equation}\label{marion:negative-milstein}
\mathrm{Prob}(x_{n+1}{<}0\,|\,x_n{>}0) =
\frac{1}{2} \left(
\mathrm{Erf}\left(\frac{\sqrt{\Delta}-1}{2\sqrt{h}\,\sigma x_n}\right) +
\mathrm{Erf}\left(\frac{\sqrt{\Delta}+1}{2\sqrt{h}\,\sigma x_n}\right)
\right),
\end{equation}

\noindent where

\begin{equation}\label{marion:Delta}
\Delta = 4h(\sigma^2x_n^2 - rx_n-x_n)
\end{equation}
\endnumparts

\noindent provided that $\Delta>0$; otherwise $x_{n+1}$ cannot become negative.
This last condition means that if

\begin{equation}\label{marion:condition}
x_n \leqslant x_{\mathrm{min}}=
\frac{r\sqrt{h} + \sqrt{3\sigma^2+hr(r+4\sigma^2)}}{2\sqrt{h}\,\sigma^2}
\end{equation}

\begin{figure}
\begin{center}
\includegraphics{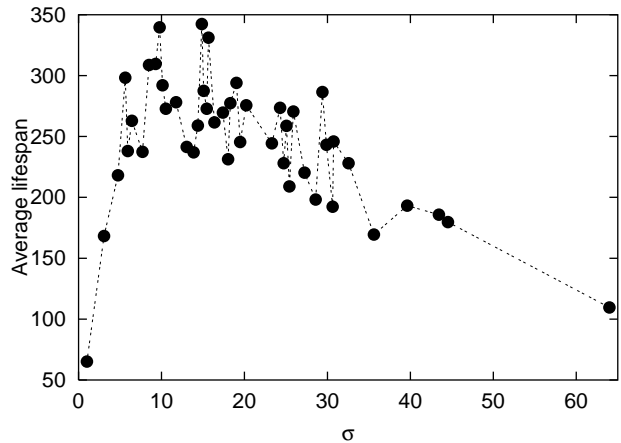}
\end{center}
\caption{Same as Fig.~\ref{marion:fig-m1} but for the Milstein
scheme (\ref{marion:Milstein}). The lifespan and $\sigma$ are in the same
units as on Fig.~\ref{marion:fig-m1}.}
\label{marion:fig-m2}
\end{figure}

\noindent the next iterate must be positive. For a small time step or for
a small noise intensity, $x_{\mathrm{min}}$ can be large. This is good news as
this prevents the discretized system from becoming unphysical. On the other hand,
in the limit of a very strong noise, $\mathrm{Prob}(x_{n+1}{<}0\,|\,x_n{>}0)\to
\mathrm{Erf}(1)\simeq0.843$ which is much larger than the corresponding value for
the Euler-Maruyama scheme. The same happens if $x_n$ becomes large which is likely
if the noise intensity is small. There is a trade off between preventing the
system from becoming unphysical for small values of the population and a rapid
divergence once the populations builds up sufficiently large. As values of $x_n$
are moderate for most of the time, we expect that the Milstein scheme
would lead to larger lifespans than Euler-Maruyama, but the qualitative
picture is much the same as in the latter: the lifespans for both small and
large noise intensities are small and there is a range of noise intensities
that maximize the average lifespan. These predictions have been confirmed
numerically. Simulations were run under the same conditions as in case of the
Euler-Maruyama scheme. Results are presented on Figure~\ref{marion:fig-m2}.
Indeed, the average lifespans first increase, then reach a range of elevated
values, and eventually decrease.
Note that the maximal lifespans are more than two times larger than those
for the Euler-Maruyama scheme. The structure seen on Figure~\ref{marion:fig-m2}
(there are apparently two maxima visible) does not necessarily correspond to any
realistic properties of the iteration (\ref{marion:Milstein})
applied to the equation (\ref{marion:eksplozja}). Rather than that, this structure may 
reflect a very pronounced stochastic variability of the system: With the
Milstein scheme, an \textit{average} lifespan of the order of $2.5\times10^2$
corresponds to a pool of runs with lifespans ranging from very short
($\sim10^{-3}$) to very large ($>10^4$) values.

\section{Conclusions}\label{conclusions}

It is well known that a discretization can dramatically change properties of a
dynamical system. The deterministic logistic equation is one classic example: its
continuous version is perfectly regular but after the discretization, the logistic
equation may display dynamical chaos. Here we have a system --- incidentally, closely
related to the logistic equation --- which, when modelled via the continuous time
SDE, is mathematically proven to remain positive and bounded almost surely, but
after a discretization it has a finite lifespan after which it becomes negative
or diverges. We have argued that, for a given time step, there exists a noise level
that is ``optimal'' in that it maximizes the average lifespan of the discretized system.
Our numerical results strongly support this argument.

We stress that the stabilization effect is
observed only in the Ito interpretation. It is usually assumed that it is the
physical interpretation of the stochastic force that should decide which interpretation
of the noise, Ito or Stratonovich, should be used~\cite{vankampen}. One may argue,
however, that a robust noise-induced stabilization should be oblivious to the noise
interpretation. With this respect, Reference~\cite{marion} and the present paper
provide an example of a system in which the consequences of the two approaches differ.

In this paper we are lucky to numerically simulate a system whose analytical 
properties are fully known. In practice, it is systems whose analytical properties
are not known that are examined by numerical simulations. If the simulations
produce divergent results, one usually concludes that the lifespan of the
system under consideration is finite. We have shown that such divergences
can be merely an artefact of a discretization of a~SDE.

Strangely, the above argument can also be reversed:
A continuous time SDE is a~mathematical idealization of a system that is discrete
by its very nature. First, populations consist of discrete individuals.
A~fractional $x$ can be interpreted only as a~fraction of a certain reference 
population. With $x\ll1$, this interpretation breaks and so may break a~model
based on a~continuous SDE. Second, it is assumed that the stochastic force
acts continually.
This idealization is fully justified when the characteristic time 
scale of the elementary random events is many orders of magnitude smaller that the
characteristic time scale of a~macroscopic process, which is the case for
example in the classical diffusion. However, when an SDE is used to describe
a biological or ecological process, the separation of time scales is less
perfect. For instance, in a biological process the characteristic time
of ``random'' changes in environmental conditions can be of the order of seconds,
and the characteristic time of the macroscopic process, like changes in a population,
can be of the order of days. In such a case the characteristic time scales are
separated by only five orders of magnitude, much as in the numerical examples 
reported above. As a~result, a process that is deterministically divergent may
remain so even after a~stochastic perturbation is applied, even though solutions
to a corresponding continuous time SDE remain bounded almost surely.

\end{document}